\documentclass[prb,twocolumn,showpacs]{revtex4}

\usepackage{graphicx}
\usepackage{amsmath}
 \usepackage{amssymb}
\usepackage{amsfonts}
\usepackage{upgreek}
\usepackage{comment}
\usepackage{color}

\renewcommand{\u}{\uparrow}
\renewcommand{\d}{\downarrow}

\begin{document}

\title{FFLO strange metal and quantum criticality in two dimensions:\\
theory and application to organic superconductors}
 
 %:\\ Theory and Application to $\kappa-$(BEDT-TTF)$_2$Cu(NCS)$_2$}

\author{Francesco Piazza$^{1,2}$}
\email{francesco.piazza@ph.tum.de}
%\homepage{http://users.ph.tum.de/ga57loj/francescopiazza/}

\author{Wilhelm Zwerger$^{2}$}
\email{zwerger@tum.de}
%\homepage{http://einrichtungen.ph.tum.de/T34/}

\author{Philipp Strack$^{3,4}$}
\email{strack@thp.uni-koeln.de}
% \homepage{http://users.physics.harvard.edu/~pstrack}

\affiliation{$^{1}$\small Institut f\"ur Theoretische Physik, Universit{\"a}t Innsbruck, A-6020~Innsbruck, Austria}
\affiliation{$^{2}$\small Physik Department, Technische Universit\"at M\"unchen, 85747 Garching, Germany}
\affiliation{$^{3}$\small Institut f\"ur Theoretische Physik, Universit\"at zu K\"oln, D-50937 Cologne, Germany}
\affiliation{$^{4}$\small Department of Physics, Harvard University, Cambridge MA 02138}

\begin{abstract}
Increasing the spin imbalance in superconductors can spatially 
modulate the gap by forming Cooper pairs with finite momentum. For large imbalances 
compared to the Fermi energy, the inhomogeneous FFLO superconductor ultimately 
becomes a normal metal. There is mounting experimental evidence 
for this scenario in 2D organic superconductors in large 
in-plane magnetic fields; this is complemented by ongoing efforts to realize 
this scenario in coupled tubes of atomic Fermi gases with spin imbalance.
Yet, a theory for the phase transition from a metal to an FFLO superconductor has 
not been developed so far and the universality class has remained unknown. 
Here we propose and analyze a spin imbalance driven quantum critical point between 
a 2D metal and an FFLO phase in anisotropic electron systems. We derive the effective 
action for electrons and bosonic FFLO pairs at this quantum phase transition. Using this 
action, we predict non-Fermi liquid behavior and the absence of quasi-particles at a discrete 
set of hot spots on the Fermi surfaces. This results in strange power-laws in thermodynamics 
and response functions, which are testable with existing experimental set-ups on 2D organic 
superconductors and may also serve as signatures of the elusive FFLO phase itself. The 
proposed universality class is distinct from previously known quantum critical metals and, 
because its critical fluctuations appear already in the pairing channel, a promising candidate 
for naked metallic quantum criticality over extended temperature ranges.

\end{abstract}


\maketitle

\section{Introduction}

%\textbf{\textit{Introduction--}} 
The concept of electronic ``quasiparticles'' moving through a 
crystal and scattering off each other at rates smaller than their 
typical kinetic energy has --together with fermionic quantum statistics-- 
led to a satisfactory understanding of the electrical conductance 
and the formation of superconductivity observed in many metals 
at low temperatures. However, an increasing number 
of observations in technologically relevant compounds 
indicate intriguing ``strange metal'' phases \cite{sachdev13}, with 
long-ranged quantum entanglement \cite{haldane15}, where
interactions destroy the electronic quasiparticles, and where electrical currents 
flow at anomalously slow rates \cite{mackenzie13}. 
\begin{figure}[t]
\vspace{0mm}
\includegraphics[width=65mm]{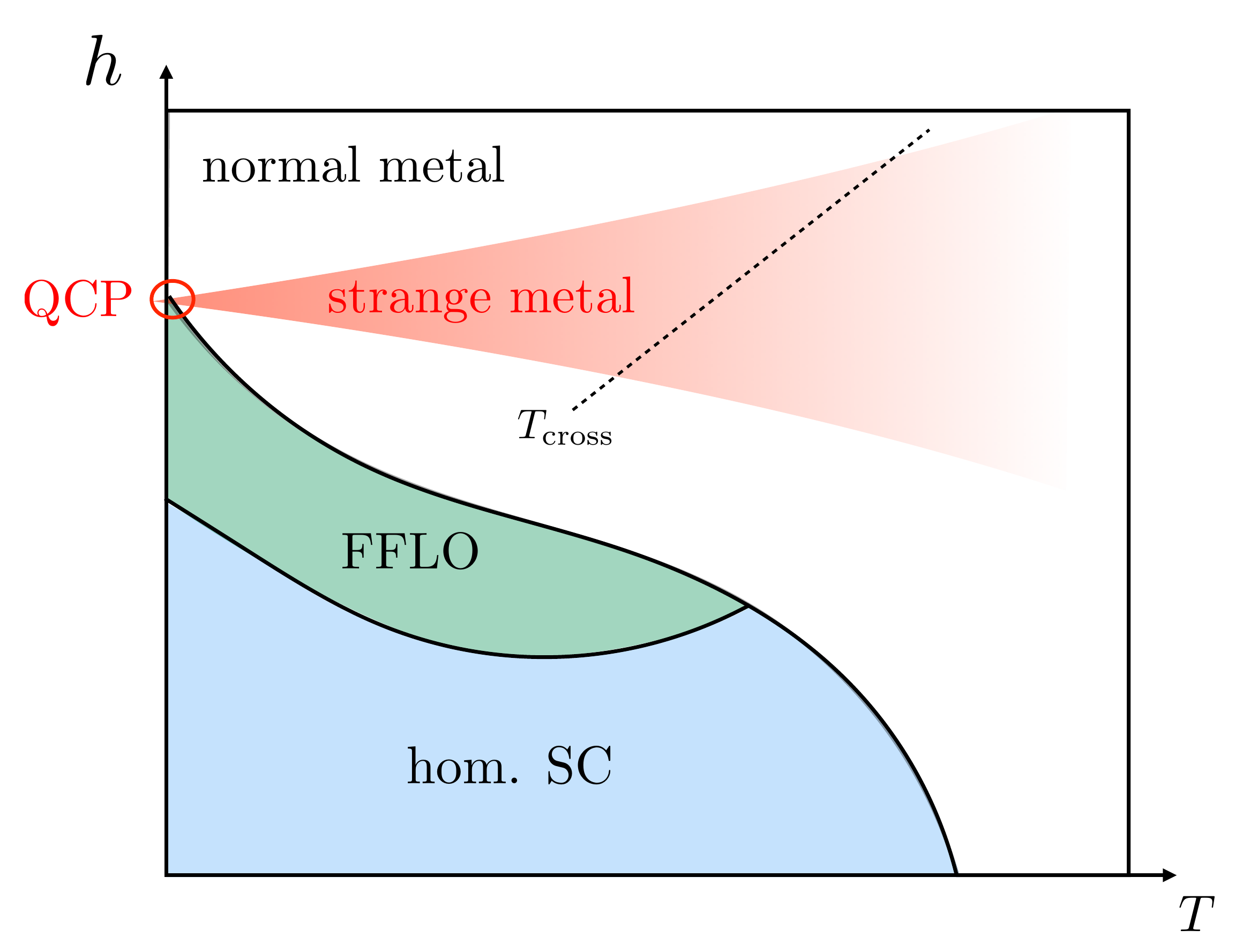}
\caption{
New strange metal phase with non-Fermi liquid behavior (red region) at the 
quantum critical point for the onset of inhomogeneous 
FFLO superconductivity in anisotropic electronic systems in an in-plane 
magnetic field ($h$) in two spatial dimensions. In the organic conductor 
$\kappa-$(BEDT-TTF)$_2$Cu(NCS)$_2$, recent evidence
places the onset of FFLO correlations (green region) to high in-plane magnetic fields $h \sim 24 - 30$T 
and low temperatures ($T < 4$K) \cite{lortz07,bergk11,mayaffre14}.
% suppression of the electronic lifetime at low temperatures as 
%$\Gamma_{\rm FFLO}^{-1} \sim \omega^{2/3}$. The absence of quasiparticles over regions 
%of the Fermi surface is 
%consistent with non-Fermi liquid features in the specific heat measurements \cite{lortz07,mayaffre14} 
%$C_{e}/T \sim T^{{1/z_f}-1}$ in the strange metal phase 
%at $h=24, 28$ Tesla with the universal 
%dynamical exponent $ z_f = 3/2$ provided by our theory. 
The
black-dashed line is 
crossover scale $T_{\rm cross}$ explained in the text.
% $\frac{t_y}{\epsilon_{\rm F}}h$ below which the physics 
%is dominated by the immediate vicinity of parts of 
%Fermi surface (hot-spots) connected by $\mathbf{Q}_{\rm FFLO}$. Here
%$t_y$ is the hopping tangential to the Fermi surface at the hot-spot
%and $\epsilon_{\rm F}$ the Fermi energy.
}
\label{phase_diagram_overall}
\end{figure}

Here, we propose a new strange metal phase associated with an 
underlying quantum critical point (red region in Fig.~\ref{phase_diagram_overall}) 
in anisotropic electronic systems
at the onset of inhomogeneous
FFLO (Fulde-Ferrell-Larkin-Ovchinikov) superconductivity 
\cite{fulde64,larkin65,casalbuoni04} in two spatial dimensions. We argue that  
organic superconductors \cite{ishiguro08,lebed08} 
are promising candidates hosting this new phase of matter in 
which electronic quasiparticles are destroyed over regions of the Fermi surface from 
scattering off FFLO waves carrying a finite momentum. As a consequence, specific 
heat and NMR (nuclear magnetic resonance) relaxation rates become non-Fermi liquid  
power-laws, whose exponents we compute. We propose that a new round of data-taking 
with existing experimental set-ups \cite{lortz07,mayaffre14} on for example 
the compound $\kappa-$(BEDT-TTF)$_2$Cu(NCS)$_2$ could provide a 
near-future experimental test of our predictions.

\subsection{FFLO phase in organic superconductors}
%{\textbf{\textit{FFLO phase in organic superconductors--}} 
Organic superconductors have emerged as leading 
candidates for observing FFLO superconducting states in electronic systems 
\cite{lortz07, shinagawa07,yonezawa08, bergk11, brown11,beyer12,mayaffre14}; 
complementary efforts are undertaken in heavy Fermion
\cite{pfleiderer09} and iron-based \cite{cho11,ptok13} compounds,
as well as in ultracold atoms \cite{zwierlein06,hulet10,thomas15}.
The crystals of organic superconductors are grown relatively clean, with a 
mean-free path larger than the coherence length of the pairs carrying a 
finite momentum $\mathbf{Q}_{\rm FFLO}$. 
For in-plane magnetic fields, superconductivity can be destroyed by a Zeeman splitting 
before orbital effects become relevant. 
%As shown by Chandrasekhar and Clogston, 
For an isotropic superconductor with no orbital effects,
the critical Pauli limit field $H_P$ is reached when the Zeeman energy $\mu_B H_{P} > \Delta/\sqrt{2}$ 
overcomes the gap~\cite{chan62, clog62}. 
A Zeeman driven transition \cite{brown11} 
requires an orbital critical field $H_{{\rm orb}}$ larger than $H_P$.
The Maki parameters $\alpha = 
\sqrt{2} H_{\rm orb}/H_{P}$ must thus be larger than one, as 
in $\kappa-$(BEDT-TTF)$_2$Cu(NCS)$_2$ where $\alpha \simeq 8$ \cite{lortz07}. 

%As noted first by Fulde and Ferrell~\cite{fulde64} and by Larkin and Ovchinnikov~\cite{larkin65},
%the transition out of a spin-singlet superconductor to a state with a finite imbalance of
%spins may appear through an intermediate phase in which a finite imbalance 
%still coexists with superconducting order. Rather than breaking up, the electrons in a 
%Cooper pair become spatially modulated by creating zeros of the superconducting order 
%parameter at which metallic quasiparticles coexist with superconductivity: the FFLO phase.

Due to only very weak and/or incoherent electron motion between the 
two-dimensional layers ($\epsilon_F/t_{\rm interlayer} \sim 3700$ for 
$\kappa-$(BEDT-TTF)$_2$Cu(NCS)$_2$ \cite{lortz07} and $t_{\rm in-plane}/t_{\rm interlayer} \sim 500$ 
in the Bechgaard salt (TMTSF)$_2$ClO$_2$ \cite{yonezawa08,lebed10}), 
out-of-plane orbital currents dissipate quickly. This is essentially true for the two compounds 
mentioned previously and also in  
the $\beta''$-phase compound $\beta''$-(BEDT-TTF)$_2$SF$_5$CH$_2$CF$_2$SO$_3$,
whose phase diagram and NMR spectra do not depend on the in-plane orientation of 
the ${\bf B}$-field \footnote{J. Wosnitza, private 
communication}.

%Upon increasing a Zeeman field $h$ in a homogeneous superconductor, it is clear 
%that the Cooper pairs would like to ``sacrifice'' as few electrons as possible to the 
%normal state: pairing is favored at momenta $\mathbf{Q}_{\rm FFLO}$ which 
%connect parts of the Zeeman-mismatched Fermi surfaces which are nested and are 
%therefore associated with a large density of states.
%in such a way that the Cooper pairs carrying momenta $\mathbf{Q}_{\rm FFLO}$  can source their electrons most effectively;
%$\mathbf{Q}_{\rm FFLO}$ prefers to connect nested parts of the Fermi surface (typically 
%called hot spots) in which the dispersion of electrons is flat in one direction and the density 
%of states is therefore enhanced. 
%In case of several nesting possibilities, the minimal 
%magnitude $|\mathbf{Q}_{\rm FFLO}|$ is preferred. 

Fig.~\ref{fig:fermisurfaces} shows three approximate Fermi surfaces from FFLO-candidate 
organic superconductors that share flat regions to which ${\bf Q}_{\rm FFLO}$ would couple to 
preferentially. ${\bf Q}_{\rm FFLO}$ are also incommensurate with 
the underlying crystal momenta. %The in-plane anisotropy of the electronic motion in the 
%organic superconductors helps single out preferential, discrete $\mathbf{Q}_{\rm FFLO}$ 
%vectors.
%
%In this paper we show that simultaneous with the unconventional 
%superconductivity in an FFLO state, there also appears a new strange metal in
%the normal phase which exhibits striking properties amenable to experimental test.
%Specifically, the phase above the QCP separating the unconventional 
%superconductor and the normal phase is a non-Fermi liquid in which 
%parts of the Fermi surface exhibit an anomalously large lifetime broadening
%$\sim \omega^{2/3}$ due to scattering off pairing 
%fluctuations associated with the incipient $\mathbf{Q}_{\rm FFLO}$ order. 
%In particular we computes the critical dynamics of the electrons in the 
%vicinity of the hot spots of the Fermi surfaces connected by $\mathbf{Q}_{\rm FFLO}$ 
%at the onset of the incommensurate FFLO superconductivity at low temperatures. 
%Specifically, our predictions for the specific heat 
%and NMR-relaxation are consistent with experimental results on $\kappa-$(BEDT-TTF)$_2$Cu(NCS)$_2$.
%
We expect our main results to be (i) universal across the various Fermi surface topologies, 
as well as (ii) relevant for both uni-directional (discussed later) and bi-directional FFLO 
modulation provided that the Fermi surface has (at least locally) nested parts which can --and will-- be connected 
by fluctuating $\mathbf{Q}_{\rm FFLO}$ order.

\section{Model} 
\begin{figure}[t]
%\hspace*{0mm}
\includegraphics[width=85mm]{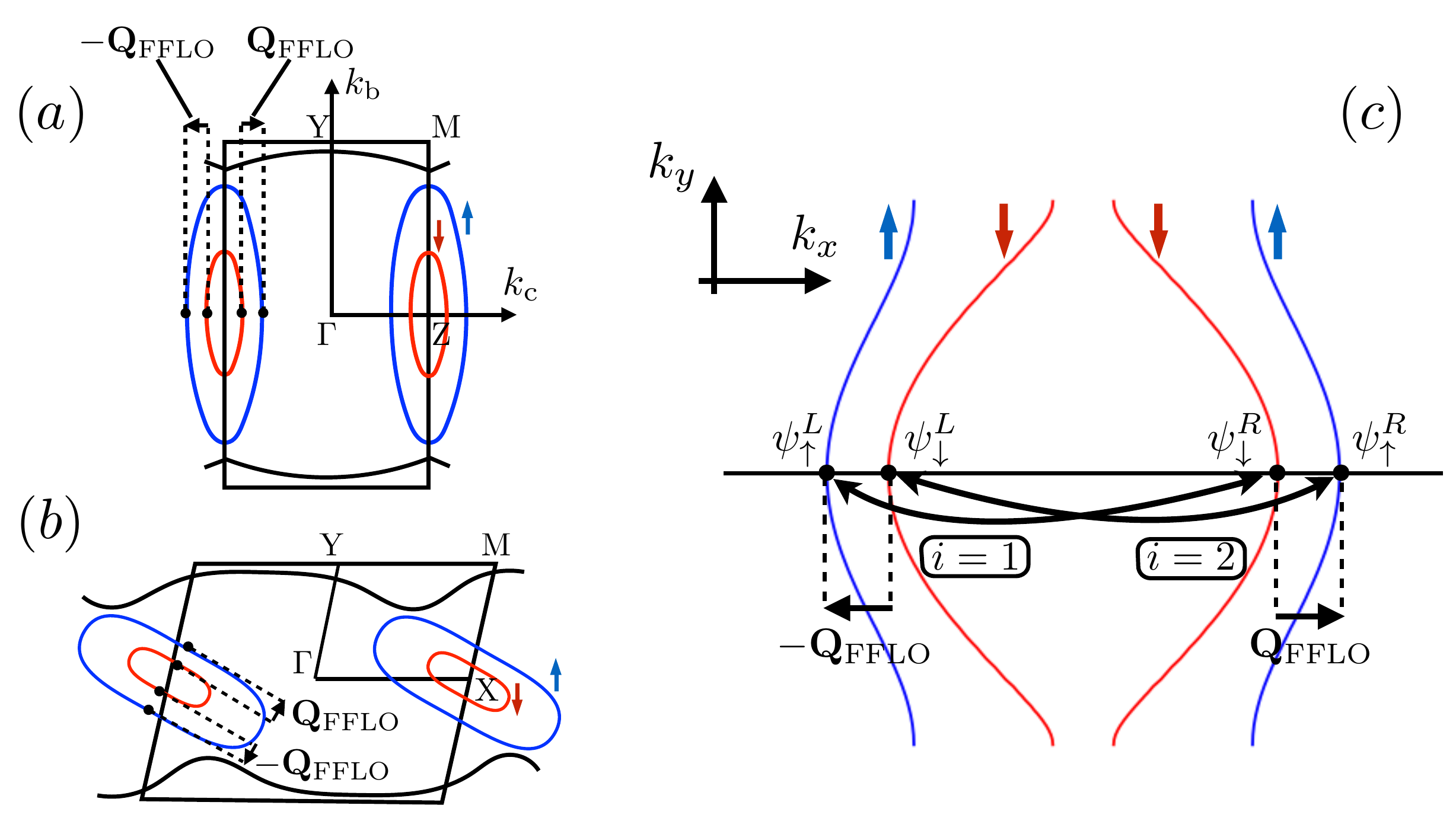}
\caption{Approximate Fermi surfaces geometry of candidate materials for the FFLO 
phase and associated strange metal. 
{\bf (a)}: $\kappa-$(BEDT-TTF)$_2$Cu(NCS)$_2$ with closed elliptic hole pockets 
corresponding to a % de-Haas-van-Alphen 
orbit frequency of about 600 Tesla \cite{lortz07}
and open parts roughly aligned with the $c$-axis \cite{wosnitza07}. Therefore an
in-plane magnetic field of $h= 30$ T results in a $\sim5\%$ mismatch
(exaggerated in the figure) of the spin-up and spin-down 
Fermi surfaces. Preferential FFLO ordering possibilities include a uni-directional modulation 
in $c$-direction 
along the short axis of the hole pocket, where $\mathbf{Q}_{\rm FFLO}$ would 
nest and connect the (Zeeman-split) sides of the hole pocket.
 {\bf (b)}: $\beta''$-(BEDT-TTF)$_2$SF$_5$CH$_2$CF$_2$SO$_3$
 with tilted hole pockets. A preferential two-dimensional 
 $\mathbf{Q}_{\rm FFLO}$, with components along the $b$ and $c$ axis,
 would again connect the weakly curved parts of the hole pocket at 
 a right angle. In both the $\kappa$- and $\beta''$-phase interlayer transport is weak 
 and incoherent such that orbital effects can be neglected.
  {\bf (c)}: Open Zeeman-split Fermi sheets characteristic for the two-dimensional 
  arrays of conducting chains of the Bechgaard salt (TMTSF)$_2$ClO$_2$ \cite{yonezawa08,lebed10} 
  in an in-plane magnetic field. 
  %The approximate hierarchy of electron hopping elements 
  %is $t_x \sim 1340$ K along the conducting chains, $t_y \sim 134$ K interchain hopping and 
  %a much weaker interlayer hopping $t_z \sim 2.6$ K. 
  The precise warping of the Fermi sheets also depends on the applied pressure 
  \cite{bourbonnais12}.}
\label{fig:fermisurfaces}
\end{figure}

\begin{figure}[t]
%\hspace*{-5mm}
\includegraphics[width=80mm]{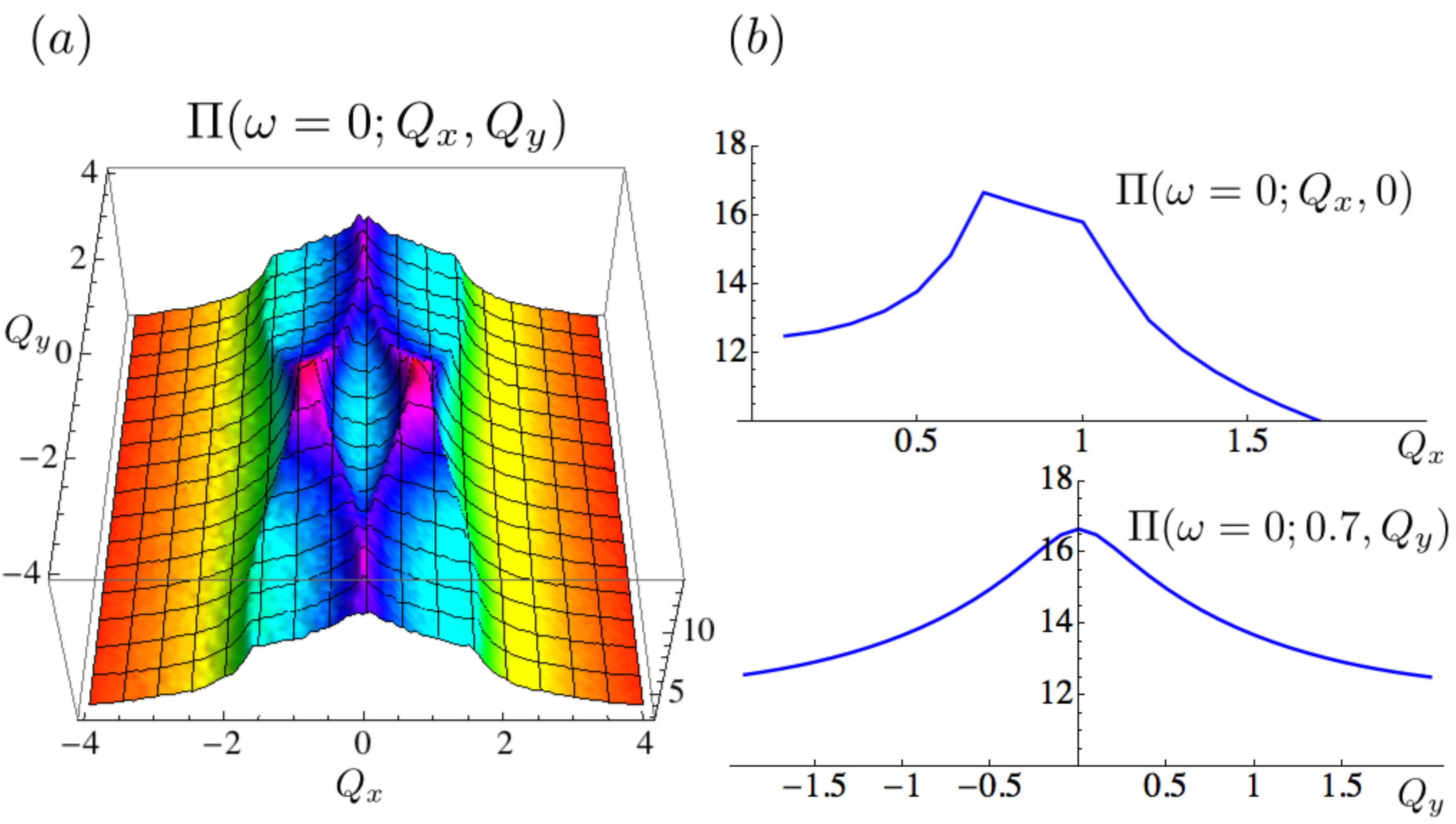}
\caption{
{\bf (a)}: Pairing susceptibility $\Pi(\omega_n=0,\mathbf{Q})$ 
at zero frequency and finite momentum for the Fermi surfaces c) in
Fig.~\ref{fig:fermisurfaces}. 
The position of the maximal peaks at  $\mathbf{Q}_{\rm FFLO}=(Q_0,0)$ 
determine the, here unidirectional, modulation wave vector of the FFLO-phase. 
{\bf (b)}: cuts of the pairing susceptibility around the peak at
$\mathbf{Q}_{\rm FFLO} $, displaying the asymmetrically-sloped ridges
along $Q_x$ characteristic of incommensurate symmetry-breaking. To capture 
quantum fluctuations around the peaks $\pm\mathbf{Q}_{\rm FFLO}+\delta{\mathbf{q}}$, 
we later parameterize the pairing fluctuation propagator with a linear 
slope from the ``outside'' and a non-analytic square-root from the
``inside'' ridge. Along $Q_y$ the peak is symmetric.
The susceptibility 
%\begin{align*}
%\label{ew:pairing_susc}
$\Pi(\omega_n,\mathbf{q})
=\frac{g^2}{\beta}\sum_{\mathbf{k}}
\frac{1-n_F(\beta\xi_\d(\mathbf{q-k}))-n_F(\beta\xi_\u(\mathbf{k}))}{\xi_\u(\mathbf{k})+\xi_\d(\mathbf{q-k})-i\omega_n}
$
%\end{align*}
involves a sum $\sum_{\mathbf{k}}$ over available states in momentum space (numerator) 
weighted by the inverse of their energies (denominator). 
Here $n_F(E)=(1+\exp(\beta E))^{-1}$ is the Fermi function and 
$\beta=1/k_B T$ the inverse temperature. 
Parameter values for the plot are 
$t_y=0.5\ \hbar^2/md^2$, $\mu=3.3\ \hbar^2/md^2$, $h=1.0\ \hbar^2/md^2$. 
}
\label{fig:fermi_surface_pairing}
\end{figure}

% for anisotropic electron systems}
%
%\textbf{\textit{Effective model--}}
%
To understand the main features of prototypical anisotropic 2D Fermi surfaces  
depicted in Fig.~\ref{fig:fermisurfaces}, let us consider a simplified model with
a spin-dependent single particle dispersion 
\begin{align}
\xi_\sigma(\mathbf{k})=k_x^2/2m-2t_y\cos(dk_y)-\mu-\sigma h.
\label{eq:dispersion}
\end{align}
Here, $m$ is the effective mass for the motion along the chains in $x$-direction. The chains are coupled by a 
weak hopping matrix element $t_y$ as is the 
case for example in the Bechgaard salt (TMTSF)$_2$ClO$_2$\cite{yonezawa08,lebed10}.
There, the hopping parallel to chains is $t_x \sim 1340$ K, between the chains 
$t_y \sim 134$ K and much weaker in the out-of-plane direction $t_z\sim 2.6$ K 
\cite{lebed10}, the latter is therefore neglected in $\xi_\sigma(\mathbf{k})$. 
Orbital effects from $t_z$-hopping influence the (material-specific and non-universal) location of $h_{c2}$ \cite{dupuis94,yonezawa08,lebed10}. 
For the $\kappa$-compound the interlayer motion is very small and for
the $\beta''$-compounds even incoherent, and plays therefore no role \cite{wosnitza07}. 
$d$ is the distance between the chains ($\sim7$\AA\, in (TMTSF)$_2$ClO$_2$)), 
$\mu$ is the chemical potential, and $h$ the Zeeman field which splits the energies 
for spin-up ($\sigma = 1$) and spin-down electrons ($\sigma = -1$).
%Our model and dispersion is applicable to in-plane magnetic fields.
%Tilted out-of-plane magnetic fields and larger coherent interlayer 
%electron motion
%can lead to an interesting interplay of FFLO formation with vortex dynamics 
%\cite{dupuis94,bergk11}. 
The key feature of $\xi_\sigma(\mathbf{k})$ is
a nested pair of so called hot spots (indicated by black points in
Fig.~\ref{fig:fermisurfaces}) connected by $\mathbf{Q}_{\rm FFLO}$; the universal feature it shares 
with the Fermi surface topologies also of $\kappa-$(BEDT-TTF)$_2$Cu(NCS)$_2$
for short-axis FFLO modulation, and 
$\beta''$-(BEDT-TTF)$_2$SF$_5$CH$_2$CF$_2$SO$_3$with diagonal 
two-dimensional FFLO modulation as per Fig.~\ref{fig:fermisurfaces}.
%\begin{widetext}
%\end{widetext}
%
The most relevant interaction for the onset of spin-singlet FFLO pairing
is a short-range attraction $g$:
\begin{align}
\label{eq:attraction}
\hat{H}_{\rm int}=-g\int d^2\mathbf{r}\;\hat{\psi}^\dag_{\u}(\mathbf{r}) 
\hat{\psi}^\dag_{\d}(\mathbf{r}) \hat{\psi}^{\phantom{\dag}}_{\d}(\mathbf{r}) 
\hat{\psi}^{\phantom{\dag}}_{\u}(\mathbf{r})\;,
\end{align}
where the electrons are represented 
by anti-commuting operators $\hat{\psi}_{\sigma}^\dag$, $\hat{\psi}_{\sigma}$.
For our calculations below, we can assume a weak coupling scenario, where
the microscopic origin of the pairing interaction plays no role
% if energies are measured in terms of the resulting gap parameter $\Delta$.
\footnote{Microscopically, the attractive interactions may thus be either due to 
a phonon-exchange mechanism or due to spin-fluctuation mediated pairing.}
\footnote{Note also that a model with the simple single particle
  dispersion $\xi_\sigma(\mathbf{k})$ of Fig.~\ref{fig:fermisurfaces}(c) and an adjustable attraction can be realized with ultracold atoms, where both the band structure and strength of the attraction can be engineered 
via optical lattices and Feshbach resonances \cite{parish07,laeuchli08,fhm09,hulet10,lutchyn11,thomas15}.}.
The Fermi surface mismatch between spin-up $\uparrow$ 
and spin-down $\downarrow$ electrons 
frustrates conventional homogeneous BCS pairing, where 
the Cooper pairs 
$
\Delta_{\rm BCS} = 
\langle
\hat{\psi}_{-\mathbf{k},\uparrow} 
\hat{\psi}_{ \mathbf{k},\downarrow}
\rangle
$
carry zero net momentum and the superconducting gap 
function is homogeneous in space. As analyzed in
Fig.~\ref{fig:fermi_surface_pairing} for the specific
$\xi_\sigma(\mathbf{k})$ given above, 
the mismatched Fermi surfaces force the electrons to pair at a finite momentum $\mathbf{Q}_{\rm FFLO} = (Q_0,0)$ 
leading, here in $x$-direction along the chains, to a modulated 
order $\Delta_{{\rm FFLO}} =\Delta_0 \cos(\mathbf{Q}_{\rm FFLO}\cdot\mathbf{r})$, 
like at a liquid crystal transition \cite{leo09,samokhin11}.

%\section{Quantum field theory}
% for the FFLO strange metal}
%
%
\section{Novel effective action}
%\textbf{\textit{Novel effective action--}}
At zero temperature $T=0$, the FFLO state forms out 
of the imbalanced metal at large magnetic fields upon decreasing $h$ 
via a continuous quantum phase transition at $h=h_{\rm QCP}$ in 
Fig.~\ref{phase_diagram_overall} (see also Appendix). 
Gaps open 
at the hot spots, i.e. at the discrete points on the Fermi surface connected 
by $\mathbf{Q}_{\rm FFLO}$.
Our aim is now to describe quantum fluctuations of electrons around the hot spots 
that can scatter off the incipient FFLO order with 
momentum transfer $\pm\mathbf{Q}_{\rm FFLO}+\delta{\mathbf{q}}$ between hot spots.
%We want to predict the anomalous thermodynamic scaling  of the FFLO strange metal 
%in the red, quantum-critical fan region in Fig.~\ref{phase_diagram_overall}. These 
%scaling laws also serve as independent evidence for the FFLO phase.
We parameterize the electron operator $\hat{\psi}_{\u\d}$ by 
fermionic quantum fields $\psi_{\u\d}^{R,L}(\tau,\mathbf{k})$ corresponding to 
the right and left part of the Fermi surface, as indicated in 
Fig.~\ref{fig:fermisurfaces} (c). The construction goes completely analogously for 
the Fermi surfaces Fig.~\ref{fig:fermisurfaces} (a,b).
Fluctuations of the FFLO order parameter are described by one bosonic 
complex quantum field $\Delta(\tau,\mathbf{k})_{1,2}$ per hot spot pair, with $\tau$ the (imaginary) 
time coordinate. 
Fluctuations of both, the electrons and the FFLO field, are encoded in the partition function
\begin{align}
Z=\int D\{\bar{\psi}_{\u,\d}^{L,R},\psi_{\u,\d}^{L,R}\}D\{\Delta_{1,2}^*\Delta_{1,2}\}\exp(-\mathcal{S})\;,
\end{align} 
where the exponential of the action $\mathcal{S}=\int_0^\beta d\tau\int d^2 \mathbf{r}\;\mathcal{L}$
weights the field configurations according to:
\begin{align}
\label{eq:lag_LR}
\mathcal{L}&=g\sum_{i=1,2}|\Delta_{i}|^2+\sum_{\substack{\sigma=\u,\d\\j=R,L}}
\bar{\psi}_{\sigma}^{j}\left(\partial_\tau-i
  v_{\sigma}^j\partial_x+\frac{\partial_y^2}{2m_y}\right)\psi_{\sigma}^{j}\nonumber\\
&-g\left[\left(\Delta_{1}^*\psi_\d^R \psi_\u^L+\Delta_{2}^*\psi_\d^L \psi_\u^R \right)+\text{h.c}\right]\;.
\end{align}
Here we have expanded the electronic dispersion around the hot spots  
$
\xi_\sigma^{R,L}(\mathbf{k})=v_\sigma^{R,L}k_x+k_y^2/2m_y
$, 
with $x,y$ being the direction orthogonal and parallel to the Fermi
surface, respectively. 
% We can assume that the Fermi velocity for the 
% $\uparrow$-electron has 
% the larger Fermi wavevector and its Fermi velocity is larger than 
% that of the $\downarrow$-electron, $v_\uparrow^{R/L} > v_\downarrow^{R/L}$ 
% (see also Fig.~\ref{fig:hotspot_theory}). 
$m_y$ is (inversely) proportional to the curvature of the Fermi
surface (and thus to $t_y$)
and related to the applied pressure in the Bechgaard salt 
(TMTSF)$_2$ClO$_2$\cite{yonezawa08,lebed10,bourbonnais12}. 
Further, we note that in the absence of time-reversal symmetry Eq.~(\ref{eq:lag_LR}) 
is not invariant under the spin-flip transformation $\uparrow \leftrightarrow \downarrow$.
The fermion-boson vertex $g$ converts bosons with momentum $\mathbf{Q}_{\rm FFLO}$ 
into electron pairs, one at each hot spot of a given pair. 
%To our knowledge the quantum field theory Eq.~(\ref{eq:lag_LR}) for FFLO quantum criticality 
%in two dimensions has not been derived and discussed previously.
The finite curvatures of the Fermi surface allow to restrict Eq.~(\ref{eq:lag_LR}) to 
the hot spots provided  temperatures are smaller than $T_{\rm cross}\sim t_y h/\epsilon_{\rm F}$
(cf. Fig.~\ref{phase_diagram_overall}).
\subsection{Distinct universality class}
%\textbf{\textit{Distinct universality class--}}
The electronic propagators in Eq.~(\ref{eq:lag_LR}) bear some similarity to those 
in two-patch theories for metals coupled to nematic 
and $U(1)$-gauge field fluctuations \cite{lee09,max_nematic10}
as well as for the onset of incommensurate $2 k_{\rm F}$ antiferromagnetism \cite{altshuler95}.
An important difference to these latter cases is that the 
electronic interactions in the particle-hole channel and that they
preserve time-reversal symmetry. 
In contrast, Eq.~(\ref{eq:lag_LR}) defines 
a new universality class as it describes 
singular electronic interactions in the (FFLO) pairing channel and the 
Zeeman field breaks time-reversal symmetry. 
%We emphasize 
%that the non-Fermi liquid phase found below arises from the singular 
%scattering off incipient finite momentum pairs in a weak coupling scenario, i.e.
%it does not require strong electron-electron interactions of the order 
%of the electronic bandwidth.

%\section{Impact of quantum fluctuations}
%
%{\it \bf Non-Fermi liquid behavior of electrons}
\section{Anomalous quantum dynamics}
%\textbf{\textit{Anomalous pairing dynamics--}}
%
The first step toward understanding quantum fluctuations 
is to compute the dynamics of the bosonic FFLO pairing 
field 
$D_{i= 1}(\tau,\mathbf{r})=\langle\hat{\Delta}_1(\tau,\mathbf{r})\hat{\Delta}_1^\dag(0,\mathbf{0})\rangle$, 
which it picks up from emission and absorption of electron pairs.
In frequency and momentum space, the interest lies in 
the slopes and direction dependence
of the pairing susceptibility close to the maxima in Fig.~\ref{fig:fermi_surface_pairing}.
The leading contribution to $D_{1}(\omega_n, \mathbf{k})$ is 
given by the one-loop bubble diagram evaluated in the Appendix. 
For the critical properties of interest here, it suffices to expand 
the (massless) fluctuation propagator directly at the quantum 
critical point for small frequencies $\omega_n$ and momenta $\pm\mathbf{Q}_{\rm
FFLO}+\delta{\mathbf{k}}$.
The result in the physically relevant small imbalance limit, $\delta
v=v_\u-v_\d\ll v=(v_\u+v_\d)/2$ (recall that an in-plane magnetic field 
of $30$ T leads to a $\sim 5 \%$ mismatch in spin-up and down Fermi surfaces 
in  $\kappa-$(BEDT-TTF)$_2$Cu(NCS)$_2$\cite{lortz07}) is:
\begin{align}
\label{pair_prop_si}
D_{1}^{-1}(\omega_n,\mathbf{k})&=\frac{V\sqrt{2m_y}}{4\pi v|\delta
  v/v|}
\left[2\mathrm{Re}\sqrt{-\frac{k_y^2}{2m_y}+\delta v
    k_x+\frac{\delta v}{v}i\omega_n}\right.\nonumber\\
&\left.+B \frac{k_y^2}{2m_y v}+C \frac{\delta v}{v}k_x\right]\;.
\end{align}
The key feature here is the square root, introducing a
non-analyticity; the usual poles of coherent excitations are
substituted by a branch cut in the complex plane. Physically, this 
reflects the overdamped, dissipative dynamics of the 
order parameter due to decay into electron pairs. 
The subleading corrections to the square root singularity 
proportional to $B,C$ are the first analytic corrections which
capture the asymmetry in the slopes of the peaks in 
Fig.~\ref{fig:fermi_surface_pairing}. Eq.~(\ref{pair_prop_si}) bears some resemblance 
to the one-loop results for the order parameter dynamics in the 
incommensurate antiferromagnetic and charge order case 
\cite{altshuler95,holder14}, which are singular fluctuations 
in the particle-hole channel of different physical origin. 
%As in these cases, the FFLO ordering wave
%vector is not commensurate with any lattice vector but  
%breaks a continuous translation symmetry.
%
%\begin{figure}[t]
%\includegraphics[width=8cm]{fig4.pdf}
%\caption{Elements of the low-energy quantum field theory for a single
%  hot-spot (see Eq.~\eqref{eq:lag_LR}). {\bf (a)}: Fermi surfaces for the two species with dispersion
%  relation
%  $\xi_\sigma^{R,L}(\mathbf{k})=v_\sigma^{R,L}k_x+k_y^2/2m_y$. A
%finite curvature is present and the Fermi velocities of the two species 
%are different. {\bf (b)}: Feynman diagrams at one-loop order. Upper row:
%propagator for the pairing fluctuations (computed in
%Eq.~\eqref{pair_prop_si}). Lower row: self-energy correction to the
%electron dispersion and quasi-particle lifetime (Eq.~\eqref{eq:self-energy}).}
%\label{fig:hotspot_theory}
%\end{figure}
%
\subsection{Non-Fermi liquid behavior of electrons}
%\textbf{\textit{Non-Fermi liquid behavior of electrons--}}
The peculiar dynamics Eq.~(\ref{pair_prop_si}) of the FFLO pairing 
fluctuations strongly back-acts on the electrons close to the
hot spots. This renormalizes the electron propagator, say, at the
$L\u$ hotspot,
$G_{L\u}(\tau,\mathbf{r})=\langle\hat{\psi}_{L\u}(\tau,\mathbf{r})\hat{\psi}_{L\u}^\dag(0,\mathbf{0})\rangle$
whose Fourier transform reads
$\left[G_{L\u}(\nu_n,\mathbf{k})\right]^{-1}
=i\nu_n-\xi_\u^{L}(\mathbf{k})$ -$\Sigma_{L\u}(\nu_n,\mathbf{k})$. 
%The electron self-energy is shown in Fig.~\ref{fig:hotspot_theory}. 
The inverse quasi-particle lifetime $\Gamma^{-1}$ at the hot spots 
for propagators of the form \eqref{pair_prop_si} can be computed
analytically \cite{holder14}. At the QCP, evaluation of the one-loop self-energy with the 
dynamical pairing propagator yields (see Appendix)
\begin{align}
\label{eq:self-energy}
\Gamma_{\rm FFLO}^{-1}=\mathrm{Im}\Sigma_{L\u}^{\rm ret}(\omega,\mathbf{q}_{\rm HS})
=\frac{1}{\sqrt{3}}\left(\frac{|\delta v/v||\omega|}{B}\right)^{2/3}\;,
\end{align}
exhibits non-Fermi liquid behavior with a fractional 
exponent $2/3$ due to the interactions with the singular FFLO pairing 
fluctuations (recall that in a Fermi liquid $\Gamma_{\rm FL}^{-1}\sim \omega^2$). 
The strength of the anomalous decay in Eq.~(\ref{eq:self-energy}) is proportional 
to the difference $\delta v$ of the Fermi velocities which itself is proportional to the 
applied Zeeman field.  In the language of critical phenomena Eq.~(\ref{eq:self-energy}) 
implies an anomalous frequency dimension $\eta_\tau=1/3$ as the bare propagator is 
linear in frequency \cite{abanov00,lee13,sur15}. Probably weaker non-Fermi liquid behavior may also appear in homogeneous 
Fermi systems with Zeeman-imbalance \cite{samokhin06,lamacraft10,strack14}. 
A similar computation for the electron self-energy 
in $k_x$ and $k_y$ direction yields a much weaker renormalization such
that the corresponding $\eta_k$'s vanish at the 
one-loop level. For direct loop computations (in absence of 
infrared cutoffs), 
the fermionic dynamic exponent is defined as $z^f=\frac{1-\eta_k}{1-\eta_\tau}$ and deviates from the 
Fermi liquid value ($z^f_{\rm FL}=1$) and instead attains the value
%
%\begin{align}
%\label{eq:dyn_exp}
$z^f_{\rm FFLO,\;one-loop} = \frac{3}{2}$.
%\end{align}
%
We note this value is also obtained in nematic metals at the one-loop level 
\cite{metzner03} but receives quantitative 
corrections at higher loops \cite{max_nematic10,holder15}. We expect 
diagrammatic expansions of the path integral of Eq.~(\ref{eq:lag_LR}) 
to be different to the nematic metal, however. One-loop corrections 
to the fermion-boson vertex, for example, vanish here in the FFLO case.

%We now show that the absence of electronic quasiparticles at the onset of FFLO pairing 
%in a two-dimensional metal lead to predictions for thermodynamic 
%and NMR measurements for organic superconductors (see Fig.~\ref{fig:specific_heat}).

\section{Predictions for new experiments} 
%
%\textbf{\textit{Predictions for new experiments--}}
%
The general scaling form of the electronic specific heat in the quantum critical ``fan'' 
in Fig.~\ref{phase_diagram_overall} is 
\cite{senthil08,karch15}
\begin{align}
\label{eq:specific_heat_th}
C_{e} /T \sim  T^{\frac{d-\theta}{z^f}-1} =  T^{-0.33}\;,
\end{align}
where the last equality has used $z^f_{\rm FFLO,\,one-loop} = 3/2$, 
$d=2$, and $\theta = 1$ in case hyperscaling is violated. 
Recall that in a Fermi liquid $C_{e}/T$ is independent of temperature.
For $\kappa-$(BEDT-TTF)$_2$Cu(NCS)$_2$, 
the specific heat has been measured by Lortz {\it et al.} \cite{lortz07,wosnitza13} 
over a range of B-fields including values towards the
upper phase-boundary of the FFLO phase: an upswing with decreasing $T$ 
incompatible with Fermi liquid behavior can be seen at $h=28$ T 
but new rounds of data taking at high magnetic fields are needed to resolve this.

A more recent experiment, where signatures of FFLO order 
have been detected in the compound $\kappa-$(BEDT-TTF)$_2$Cu(NCS)$_2$ 
was performed by Mayaffre {\it et al.} who have measured the 
NMR relaxation rate $1/(T_1 T)$ at 
high magnetic fields including data set at $h=22$, $27$ Tesla (see Fig.~2 of 
Ref.~\onlinecite{mayaffre14}). They associate the observed upswing at $h=22$ Tesla 
with bound states of quasiparticles at the nodes of an FFLO order parameter. 
At $h=27$ Tesla, however, $1/(T_1T)$  seems to decrease for temperatures 
smaller than 3 Kelvin. 
Our strange metal sccenario predicts a power law decrease of the form
$1/(T_1 T) \sim T^{2/3}$ which is suppressed compared to a constant Fermi liquid 
relaxation rate. New rounds of data taking over extended 
temperature and higher magnetic fields should resolve this.
%A comparison to currently available data from Ref.~\onlinecite{mayaffre14} 
%is performed in Fig~\ref{fig:specific_heat}. 
The power-law exponent in $1/(T_1 T)$  
is determined by the scaling behavior of the 
density of states 
\begin{align}
\frac{1}{T_1T}&\sim\frac{1}{T}\int d\omega N^{\rm hot}_\u(\omega) N^{\rm hot}_\d(\omega) n_F(
\omega)\left[1-n_F(\omega)\right]
%\nonumber\\
%&
%\sim T^{2/3}
%< \frac{1}{T_1 T}|_{\rm FL} = {\rm const.}\;.
\label{eq:T1T}
\end{align}
with $N^{\rm hot}_\sigma (\omega) = \int \frac{ d^2
  \mathbf{k}}{(2\pi)^2} A^{\rm hot}_\sigma(\omega,\mathbf{k})$. 
  For the critical electrons at the hot spots the associated spectral
function is expected to fullfill $\omega/T$ scaling \cite{senthil08,sachdev_book}
\begin{align}
\label{eq:A_scaling}
A^{\rm hot}_\sigma(\omega,\mathbf{k}) \sim \frac{c_0}{|\omega|^{1-\eta_\tau}}
 \mathcal{F}_\sigma\left(\frac{ c_1 \omega}{(k_x+k_y^2)^{z_f}}, \frac{\omega}{T}\right)\;,
\end{align}
where the anomalous scaling in spatial directions $x$ and $y$ can 
take different finite values but do not influence $z_f$ at the one-loop level.
The same temporal anomalous dimension $\eta_\tau = 1/3$ 
that appears in the self-energy Eq.~(\ref{eq:self-energy}) leads to
$N^{\rm hot}_\sigma (\omega)\sim\omega^{1/3}$ (see Appendix)
and thus to the exponent $2/3$ in Eq.~(\ref{eq:T1T}) for electrons from the immediate 
vicinity of the hot spots. As in the nematic metal, we expect higher-loop 
corrections to further suppress the density of states \cite{max_nematic10}.
In contrast to the nematic metal, however,
where all points on the Fermi surface can be taken to be equivalent 
\cite{max_nematic10,sachdev_book}, here in the FFLO strange metal 
a discrete pair of hot spots is singled out. 
To account for electrons from ``colder'' parts of the Fermi surface 
one should add a constant Fermi liquid 
piece to the density of states $N_\sigma(\omega) = N_{0\sigma}^{\rm cold} + N_\sigma^{\rm hot}(\omega)$
and NMR relaxation rate when comparing with experiments. 
A proper analysis of crossovers and potential contributions 
violating $\omega/T$ scaling \cite{abanov03,dellanna06}
require momentum-resolved renormalization group techniques.
%
%To allow for statistically significant experiment-theory comparison, we propose the following 
%measurements: first, a magnetic field scan 
%of $C_e/T$  to higher $h$-fields to locate the true boundary of the FFLO phase $h_{\rm QCP}$ 
%which may lay above 30 Tesla
%(i.e. where the transition peak is completely absent down to the lowest 
%attainable sub-Kelvin regime). Second, the region $(T_{\rm lowest},T_{\rm cross})$ 
%$(26$ Tesla, $h_{\rm QCP})$ needs to be scanned over ideally two or
%more decades in temperature to discriminate power-law scaling. Complementary 
%measurements of both $C_e/T$ and $1/T_1 T$ would lead to independent statistics. 
%If the crossover temperature $T_{\rm cross}$ of Fig.~\ref{phase_diagram_overall} can 
%be increased by warping the Fermi surfaces e.g. through pressure, that is advantageous 
%to cover extended temperature ranges.
%
\section{Summary}
%\textbf{\textit{Summary--}}
The FFLO strange metal phase proposed here provides a new universality class 
for the study of critical thermodynamics and transport without
electronic quasi-particles. Our results open up the possibility to 
targeted new rounds of data taking in the organic superconductors
and may facilitate access to a naked quantum critical metal 
with critical fluctuations in the pairing channel
over extended temperature ranges.

\acknowledgments

%\textbf{\textit{Acknowledgments--}}
%
%We are very grateful to Wilhelm Zwerger for insightful discussions
%on all aspects of the paper and pointing us toward the experimental 
%observables of the organic superconductors. 
We thank Jochen Wosnitza for discussions and for providing us with 
the raw specific heat data of Ref.~\onlinecite{lortz07}.
Moreover, we thank Peter Fulde, Tobias Holder, Matthias Vojta, and Tarik Yefsah 
for discussions and gratefully acknowledge useful comments on the manuscript by 
Andreas Eberlein and Walter Metzner. 
PS acknowledges support by the Leibniz prize of A. Rosch, a 
MURI grant W911NF-14-1-0003 from ARO, a NSF grant DMR-1360789, and 
the MIT-Harvard Center for Ultracold Atoms. FP acknowledges support by
the APART fellowship of the Austrian Academy of Sciences.

%\newpage
%\bibliographystyle{aipnum4-1}
%\bibliographystyle{abbrv}
\bibliographystyle{ieeetr}
\bibliography{Strange}

\appendix

\begin{widetext}

\section{Continuous mean-field FFLO transition}

We decouple the interaction term (Eq.~\eqref{eq:attraction} of the main text) in the pairing channel and write the partition function as the following functional integral
\begin{align}
Z=\int D(\bar{\psi}_{\u,\d},\psi_{\u,\d})D(\Delta^*,\Delta) e^{-S_{\rm tot}[\bar{\psi}_{\u,\d},\psi _{\u,\d},\Delta^*,\Delta]}
\end{align}
for the complex Grassmann field $\psi$ and the bosonic field $\Delta$, with the euclidean interaction action
\[
S_{\rm int}=\int_0^\beta d\tau\int d^2\mathbf{r}\left[\frac{1}{g}|\Delta(\tau,\mathbf{r})|^2-\left(\Delta^*(\tau,\mathbf{r})\psi_\d(\tau,\mathbf{r}) \psi_\u(\tau,\mathbf{r})+\Delta(\tau,\mathbf{r})\bar{\psi}_\u(\tau,\mathbf{r}) \bar{\psi}_\d(\tau,\mathbf{r})\right)\right]\;.
\]
An effective Ginzburg-Landau theory for the pairing field is obtained
by integrating out the fermionic fields. Written in Fourier space the latter reads
\begin{align}
S_{\rm eff}=\frac{\beta}{ g}\sum_{\omega_n,\mathbf{k}}|\Delta(\omega_n,\mathbf{k})|^2-\mathrm{Tr}\ln\beta\underline{G}^{-1}\;,
\end{align}
where $\mathrm{Tr}=\sum_{\nu_n,\mathbf{k}}\mathrm{tr}$, with
$\mathrm{tr}$ meaning the trace in the 2x2 spinor space, the
bosonic(fermionic) Mastubara frequency $\omega_n=2\pi
n/\beta$($\nu_n=\pi (2n+1)/\beta$), $\beta$ the inverse temperature,
and with the matrix propagator
\begin{align}
\underline{G}^{-1}(\nu_n,\mathbf{k};\nu'_n,\mathbf{k}')=\left(\begin{array}{cc} \delta_{\nu_n,\nu'_n}\delta_{\mathbf{k,k'}}(i\nu_n-\xi_\u(\mathbf{k})) & \Delta(\nu_n-\nu'_n,\mathbf{k-k'})\\ \Delta^*(\nu'_n-\nu_n,\mathbf{k'-k}) & \delta_{\nu_n,\nu'_n}\delta_{\mathbf{k,k'}}(i\nu_n+\xi_\d(-\mathbf{k}))\end{array}\right)\;,
\end{align}
where $\xi_\sigma(\mathbf{k})$ is the fermionic dispersion relation.
Now, upon expanding the logarithm in powers of $\Delta$, we can compute the
coefficients of the effective Ginzburg-Landau theory. In particular,
the second order term contains the pairing susceptibility
\begin{align}
\label{bubble}
\Pi(\omega_n,\mathbf{q})=\sum_{\mathbf{k}}\frac{1-n_F(\beta\xi_\d(\mathbf{q-k}))-n_F(\beta\xi_\u(\mathbf{k}))}{\xi_\u(\mathbf{k})+\xi_\d(\mathbf{q-k})-i\omega_n}\;.
\end{align}
Specifying now to the mean-field FFLO transition with favoured
momentum $|\mathbf{Q}_{\rm FFLO}|$ we can express the paring field as
\begin{align}
\Delta(\omega_n, \mathbf{q})
\rightarrow \Delta_0 (0,\mathbf{Q}_{\rm FFLO})=\beta\delta_{\omega_n,0} 
\left[ 
d_{+\mathbf{Q}_{\rm FFLO}} \delta_{\mathbf{q},\mathbf{Q}_{\rm FFLO}} (2\pi)^2
+
d_{-\mathbf{Q}_{\rm FFLO}} \delta_{\mathbf{q},-\mathbf{Q}_{\rm FFLO}} (2\pi)^2
\right]\;.
\label{eq:decomp}
\end{align}
The $d'$s are the amplitudes of the FFLO-order parameter, that are
chosen to be equal and real.

The quadratic Landau coefficient is thus given
by
\[
a_2 = 1/g - \Pi(\omega_n = 0, \mathbf{Q}_{\rm FFLO})\;.
\]

The quartic Landau $a_4$  coefficient results from the expansion of the logarithm
to fourth order in $\Delta$ and contains all possible
contractions of four fermionic propagators with the four
external momenta fixed at $\pm\mathbf{Q}_{\rm FFLO}$ and zero
frequency. It must be numerically calculated.
All contractions diverge when $T\rightarrow 0$, indicating the need 
to devise a renormalization group procedure for the problem at 
very low temperatures. For generic choices of parameters
\begin{align}
\lim_{T\rightarrow 0} a_4 >0
\end{align}
indicative of a continuous phase transition at least at the mean-field level.

\section{Pairing fluctuation propagator, fermion propagator and
  density of states}

In this section we compute the one-loop corrections to the pairing and
fermion propagators, leading to the Eqs.~\eqref{pair_prop_si} and \eqref{eq:self-energy} of the main text.
We use the low-energy Lagrangian from Eq.~\eqref{eq:lag_LR} of the main text, whose
Fermi surface structure and one-loop diagrams are shown in Fig.~\ref{fig:hotspot_theory}.

The inverse pairing fluctuation propagator diagrammatically shown in the upper
row of Fig.~\ref{fig:hotspot_theory}(b) contains the bare part, simply given by
the Landau coefficient $a_2$ (vanishing at the critical point), plus
the pairing susceptibility:
\begin{align}
\label{eq:bubble_LO}
\Pi_{i=1}(\omega_n,\mathbf{q})=\frac{1}{\beta}\sum_{\nu_n,\mathbf{k}}G_{R\d}^{(0)}(\nu_n,\mathbf{k}) G_{L\u}^{(0)}(\omega_n-\nu_n,\mathbf{q-k})=\frac{1}{\beta}\sum_{\nu_n,\mathbf{k}}\frac{1}{i\nu_n-\xi_\d^R(\mathbf{k})}\frac{1}{i(\omega_n-\nu_n)-\xi_\u^L(\mathbf{q-k})}\;\;\;,
\end{align}

\begin{figure}[t]
\includegraphics[width=12cm]{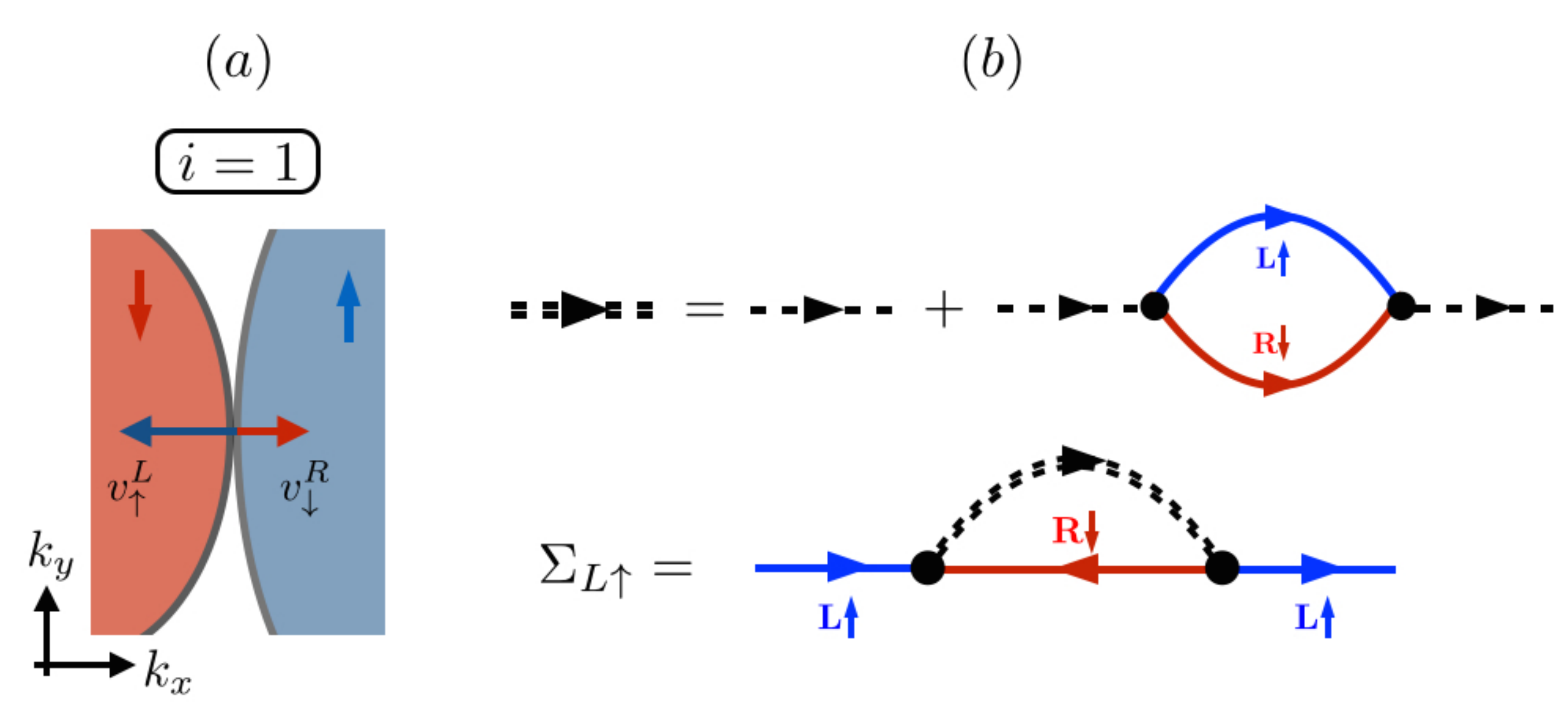}
\caption{Elements of the low-energy quantum field theory for a single
 hot-spot of the Lagrangian given in Eq.(3) of the main text. {\bf (a)}: Fermi surfaces for the two species with dispersion
 relation
 $\xi_\sigma^{R,L}(\mathbf{k})=v_\sigma^{R,L}k_x+k_y^2/2m_y$. A
finite curvature is present and the Fermi velocities of the two species 
are different. {\bf (b)}: Feynman diagrams at one-loop order. Upper row:
propagator for the pairing fluctuations, containing the susceptibility
given in Eq.~\eqref{eq:bubble_LO}. Lower row: self-energy correction to the
electron dispersion and quasi-particle lifetime, given in Eq.~\eqref{eq:se_mats}.}
\label{fig:hotspot_theory}
\end{figure}

We first perform the Matsubara sum in Eq.~\eqref{eq:bubble_LO}.
By restricting to $T=0$, we integrate first over $k_x$ and then over
$k_y$, always discarding the unphysical UV-divergent terms, and get
\begin{align}
\Pi_{i=1}(\omega_n,\mathbf{k})=&-\frac{V\sqrt{2m_y}}{2\pi(v_\u+v_\d)}\left[\frac{1}{|\frac{v_\u}{v_\d}-1|}\sqrt{-\frac{v_\u}{v_\d}\frac{k_y^2}{2m_y}+(\frac{v_\u}{v_\d}-1)(v_\u
      k_x+i \omega_n)}+\right.\nonumber\\
&\left.+\frac{1}{|\frac{v_\d}{v_\u}-1|}\sqrt{-\frac{v_\d}{v_\u}\frac{k_y^2}{2m_y}+(\frac{v_\d}{v_\u}-1)(-v_\d
      k_x+i \omega_n)}\right]\;,
\end{align}
which in the small imbalance limit $\delta v=v_\u-v_\d\ll
v=(v_\u+v_\d)/2$ gives Eq.~\eqref{pair_prop_si} of the main text since $D_{i=1}^{-1}(
\omega_n,\mathbf{k})=-\Pi_{i=1}(\omega_n,\mathbf{k})$ at the
critical point. In the main text the first analytical corrections coming from higher order terms in
the fermionic dispersion have been included in $\Pi_{i=1}(\omega_n,\mathbf{k})$.

In order to compute the non-Fermi-liquid exponent we consider the
quasi-particle decay rate given by the imaginary part of the
analytically continued self-energy. At the one-loop level the
fermionic self-energy is given by (Feynman
diagram in the lower row of Fig.~\ref{fig:hotspot_theory}(b))
\begin{align}
\label{eq:se_mats}
\Sigma_{L\u}(\nu_n,\mathbf{q})=\frac{1}{\beta}\sum_{\nu_n,\mathbf{k}}D_{i=1}(\omega_n,\mathbf{k})G_{R\d}^{(0)}(\omega_n-\nu_n,\mathbf{k-q}) \;.
\end{align}
After analytical continuing $i\nu_n\to \omega+i0^{+}$ and taking the
imaginary part we get
\begin{align}
\label{eq:self-energy-LU}
\mathrm{Im}\Sigma_{L\u}^{\rm
  ret}(\omega,\mathbf{q})=-\sum_{\mathbf{k}}\left[n_F(\xi_{R\d}(\mathbf{k}))+n_B(\xi_{R\d}(\mathbf{k})+\omega)\right]\mathrm{Im}D_{1}^{\rm
ret}(\xi_{R\d}(\mathbf{k})+\omega,\mathbf{k+q})\;.
\end{align}
We now set $T=0$, $\mathbf{q}=0$, and first integrate over $k_x$ and then over $k_y$, to get Eq.~\eqref{eq:self-energy}
of the main text. 
By computing the self-energy \eqref{eq:se_mats} at finite
momentum we obtain no $\mathbf{q}$-dependence at leading order, once
we expand the pairing propagator (Eq.~\eqref{pair_prop_si} in the main text) consistently with
the small imbalance limit, to get the Landau-damped form 
\begin{align}
\label{eq:boseprop}
D_{1}^{-1}(\omega_n,\mathbf{k})\simeq\frac{V\sqrt{2m_y}}{4\pi v|\delta
  v/v|}
\left[\sqrt{2m_y}\frac{|\delta
  v/v||\omega_n|}{|k_y|}+B \frac{k_y^2}{2m_y v}% +C \frac{\delta v}{v}k_x
\right]\;.
\end{align}
Within this approximate one-loop framework we have therefore no anomalous dimension
for the fermion momenta: $\eta_x=\eta_y=0$.

By adding the self-energy correction from Eq.~\eqref{eq:self-energy} in the main text to
the retarded fermion propagator, we can compute the spectral function
$A_\u(\omega,\mathbf{k})=-\mathrm{Im}G_{L\u}^{\rm
  ret}(\omega,\mathbf{k})/\pi$ which satisfies the scaling relation of
Eq.~\eqref{eq:A_scaling} in the main text with $\eta_\tau=1/3$. The
density of states 
\begin{align}
\label{eq:dos}
N_\u(\omega,T)\sim \omega^{1/3}
\end{align}
is then obtained by integrating the spectral function
over $k_x,k_y$ after rescaling both coordinates in order to eliminate
the $\omega$-dependence from the integrand. This procedure is based on
the idea that we have to put a UV-cutoff on both the $k_x$ and the
$k_y$ integrals since we are dealing with a low-energy theory in the
proximity of the hot-spot. 
Accordingly, we add to the hot-spot contribution \eqref{eq:dos}
(called $N_\u^{\rm hot}$ in the main text) a $\omega$-independent shift coming from the
``cold'' part of the Fermi surface (called $N_{0\u}^{\rm cold}$ in the main text), which indeed should behave like a
Fermi liquid.
However, in this way the exponent in
Eq.~\eqref{eq:dos} can be strongly overestimated. 

\end{widetext}

\end{document}